# CONFIRMATION OF SEVEN FAINT ATLAS VARIABLE STAR CANDIDATES

THOMAS, NEIL [1], GUAN, CELINA [1]

1) Department of Astronautical Engineering, United States Air Force Academy, CO 80840, USA, neil.thomas@afacademy.af.edu

**Abstract:** A pilot survey conducted at the Lookout Observatory has confirmed seven faint ($V \sim 13$ to $17$) variables in the region of Kepler-76b that were recently discovered by the Asteroid Terrestrial-Impact Last Alert System (ATLAS). The ATLAS survey identified 315,000 probably variables within its wide-field survey in 2018. The faintness (down to $r \sim 18$) and small amplitudes (down to 0.02 mag) included in these candidates makes external validation difficult. Our confirmation of seven such variable stars gives credibility to the ATLAS list. Further, the agreement between various surveys and LO data validates the use of our new survey for variable star and exoplanet research.

## 1 Introduction

The Lookout Observatory (LO) survey is primarily intended to observe exoplanets via the transit photometry method. It routinely attains photometric precision better than 0.002 mag (2 mmag) for bright targets ($V<12$). Useful photometry is generally collected down to $V\sim17$, depending on conditions and exposure length. In addition to the targeted exoplanet host star our software autonomously performs photometry on all suitable stars in the field of view (FOV). It thus also provides valuable observations of many variable stars during a given night. We present the observations of seven faint ($V=13$-$17$) variable stars previously only identified by ATLAS. We do not provide an in-depth analysis of each star but rather, we confirm their variable natures and demonstrate the quality of this survey for this type of work.

In Section 2 we discuss the characteristics and expected performance of the LO survey. In Section 3 we gain confidence in this survey by specifically comparing the results from two variables independently identified by LO which also have light curves available from other surveys. In Section 4 we confirm the existence of seven variable star candidates identified previously only by ATLAS. Section 5 is the conclusion.

## 2 Instrumentation and Methodology

The Lookout Observatory consist of an 11" Celestron telescope modified to f/1.9 with a HyperStar lens replacing the secondary mirror. Imaging is done with a ZWO ASI 1600 CMOS camera. The FOV is 114' x 86' or 2.7 deg². This typically allows for the collection of photometry on 500 to 15,000 stars depending on the region of the sky and observing conditions with a cadence of 10 to 60 seconds. Optical filters are not usually used. A single FOV is generally observed for an entire night.

LO was constructed by faculty and students of the Astronautical Engineering Department at the US Air Force Academy and emphasizes consumer grade instruments (entire observatory less than $10,000 USD). It first successfully observed an exoplanet transit in May of 2019. Although it is at an altitude of 7000', the site suffers from suburban light pollution and variable weather. About 20% of nights have long enough periods of clarity to allow for observations. Images are collected semi-autonomously using MAXIM DL and CCDCOMMANDER.

Our team developed a custom data pipeline in MATLAB to mitigate poor sky conditions. The software is especially designed to lessen the impact of variable weather and light pollution. The site has a light pollution Bortle index of four, with sky brightness being measured at 20 mag per arcsecond on moonless nights. The software also deals with issues common to small observatories such as errors in tracking, focus, and mirror flop. It allows for highly automated processing of all reasonable stars without user input for targeting, calibration star selection, or photometric aperture determination. In short, LO provides high-cadence, nearly continuous photometry for up to 15,000 stars over the time span of a night.

The LO is specifically designed to maintain the photometric precision necessary to observe exoplanet transits. Even large, short-period exoplanets generally only cause a stellar dimming of 30 mmag or less and have a duration of two to three hours. To decisively capture a transit (and a sufficient baseline before and afterwards) our goal is to maintain noise levels (photometric RMS) of less than 2 mmag from dusk to dawn for 10th magnitude stars using one-minute exposures. But useful photometry is generally collected down to $V<17$, depending on conditions and exposure length. Approximately 70 exoplanet transits and many more variable stars have successfully been observed over the past two years. The relative quality of our photometry can best be judged using Fig. 1. The standard deviation of calibrated photometry for all stars is calculated and plotted against magnitude as dots. Since most stars are stable, our intrinsic error is seen to be well below 10 mmag at $V=12$ and is consistently below 200 mmag at $V=17$. An attempt is also made to classify noise as either photon shot, scintillation, or systematic (red).

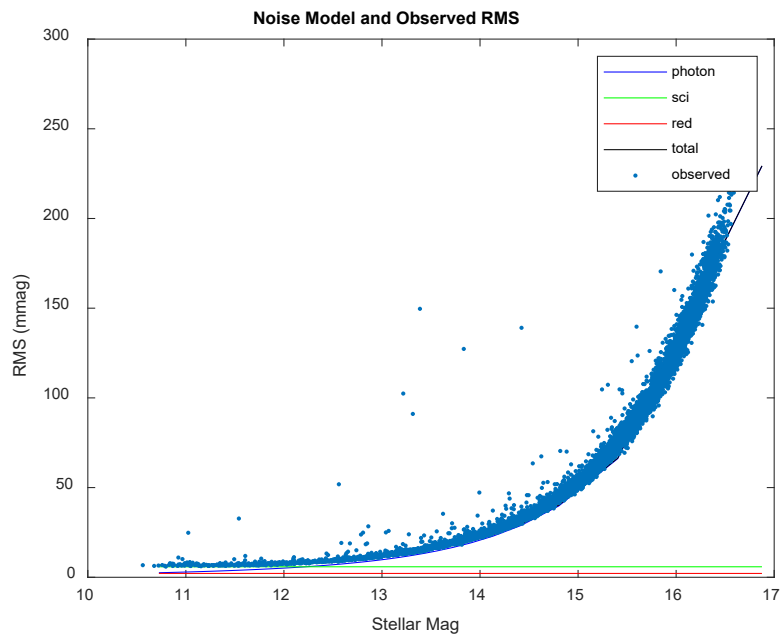

Figure 1: The magnitude of each star plotted against its photometric RMS. Except for variable star outliers, most of observed stars in the survey create a reliable correlation in agreement with photon shot noise and lesser sources of noise, such as atmospheric scintillation and red noise.

　　　We observed a FOV centered on Kepler-76b (RA 294.1921°, DEC 39.619° J2000) on five nights between September 2019 and July 2020 using 50 second unfiltered exposures. The average number of stars providing usable photometry was 14,410. The data pipeline flagged an average of 0.47% of stars as variables. Approximately 30-100 comparison stars are automatically selected for each star during its photometric calibration based on similarity in brightness and color. A typical image is shown in Fig. 2 with the target and comparison stars marked. This image is for the calibration of the host star for Kepler-76b, which is in the center. Each star, however, will have its unique set of comparison stars. A magnified view of this target star is shown in Fig. 3 to better demonstrate the FOV.

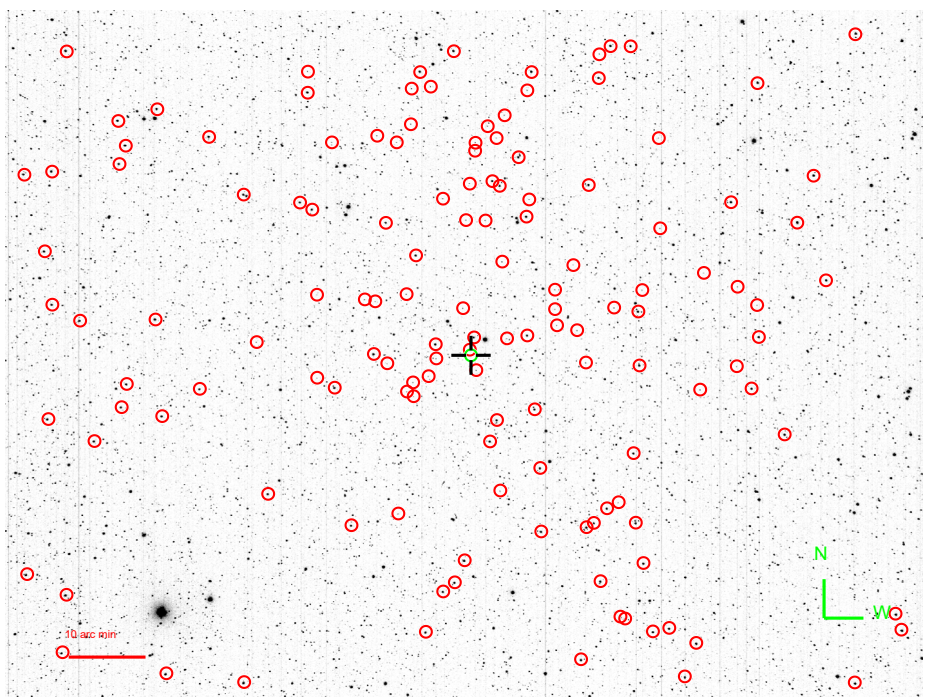

Figure 2: A sample image showing the FOV with identification of stars. This frame is reference to the host star of the exoplanet Kepler-76b, shown with crosshairs in the center. Automatically selected comparison stars are circled. The size of the circles are arbitrary in this figure.

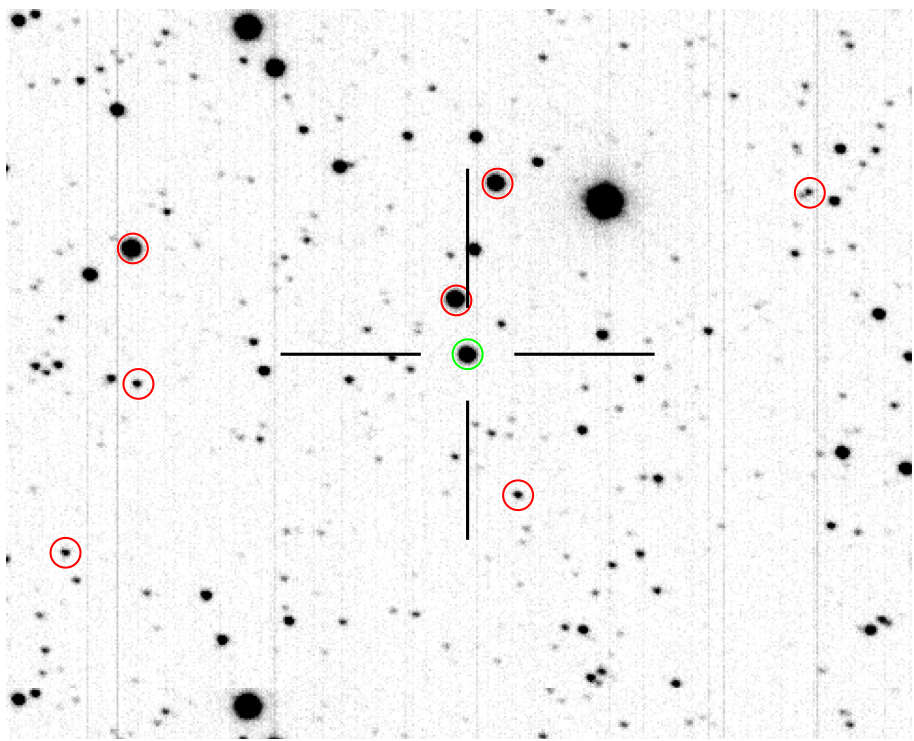

Figure 3: A magnified view of the region around this target star. Comparison stars used for calibration are now circled with a radius corresponding to their photometric apertures.

Calibrating with comparison stars provides relative magnitudes. We then use Gaia DR2 magnitudes to derive the absolute magnitudes of stars in our survey (Gaia Colab. et al. 2016, 2018). To do this, the instrumental magnitude of each star from a night is compared to its Gaia magnitude and used to determine the global shift to true magnitudes. Then a color correction is applied after fitting magnitude errors to Gaia *B-V* colors. Although our photometry is unfiltered and Gaia is *G*-band, a reliable transformation is possible. Our calibrated magnitudes are compared to Gaia values for a typical night in Fig. 4. Although noise is photon dominated for faint stars, the standard deviation of the magnitude differences for the brightest third of stars is 23 mmag. For comparison, the quoted errors of Gaia magnitudes in this brightness range are typically 2-6 mmag. The post-calibration magnitude error for each star is empirically estimated by computing the standard deviation of the observed errors seen in the ~100 stars which are closest in magnitude to a given star.

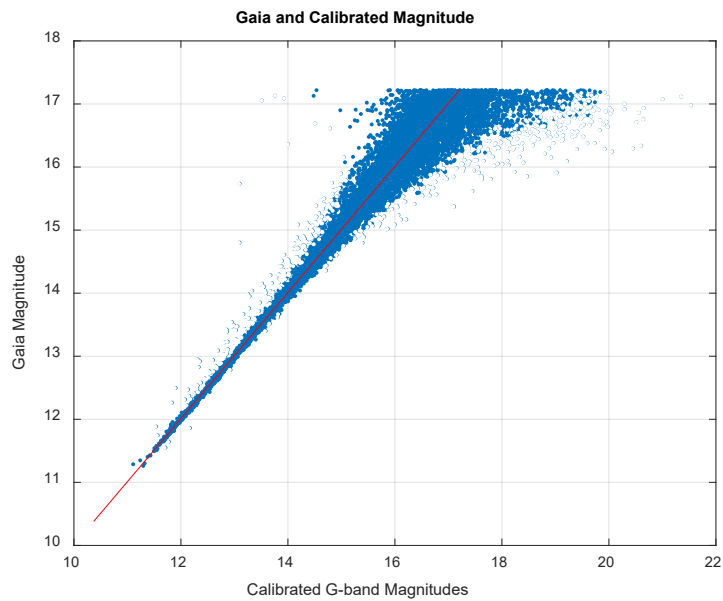

Figure 4: The results after calibrating our instrumental magnitudes to Gaia G-band. Errors in faint stars are signal limited. The brightest third of stars during this session, however, are calibrated to a difference from Gaia values having a standard deviation of 23 mmag.

Potential variables are found by calculating the Lomb-Scargle periodogram of each star's photometry and flagging those with a peak power above some threshold. A simple sinusoid fit is found for these stars. Then the standard deviation of the difference between the data and the fit is compared to the unfit standard deviation. If the improvement is great enough then the star is sent to a more advanced analysis.

Cyclic variables such as delta Scuti, ellipsoidal, and contact binaries are fit to a fourth order trigonometric polynomial using data from all available observing sessions. This provides the period of the variable and its error is derived from the covariance matrix. These errors are typically quite small due their short periods (<1 day) and data that spans the better part of a year. Amplitude is not a direct parameter of the polynomial fit. To determine amplitude, we use the polynomial to calculate magnitude values over one period and subtract the faintest value from the brightest. To determine the error in these amplitude values we employ a Monte Carlo approach in which the polynomial coefficients are varied randomly according to their own error estimates from the covariance matrix. The amplitude of each perturbed function is recorded and after 1,000 iterations the standard deviation of these amplitudes is recorded. The magnitude of the star is determined using a weighted mean of the magnitudes as found earlier. Its error is based on the errors determined empirically as discussed previously. Although the covariance matrix from the polynomial fit provides a statistical estimate of magnitude error, it should not be used. The photometry is more precise than it is accurate. The absolute magnitudes differ between nights due to slight systematic offsets. This variability is manageable but is significantly greater than the scatter within the data during a typical night. The polynomial fit is likely to be very precise and the covariance derived error will be misleadingly small because these offsets are not visible to the fitting algorithm. Hence, we rely on the empirical error estimates. Parameters for transient variables are found similarly except that a tenth order polynomial is used and the magnitude is the brightest (out-of-transit) value instead of the mean.

Seven of the stars we identified as variable were unknown as such in any source except ATLAS. The parameters as determined by our data pipeline are listed in Tab. 1. The corresponding characteristics from ATLAS are provided in Tab. 2. The classification terminology used by ATLAS is detailed in Tab. 3.

Table 1. Stellar properties derived from this survey.

| Name | $G$ (error) | Period in days (error x $10^{-6}$) | Amplitude in mmag (err) | Epoch in HJD (error) |
|---|---|---|---|---|
| KID 05015926 | 12.940 (0.013) | 0.3626882 (0.71) | 117 (2) | 2459032.8470 (0.008) |
| V1130 Cyg | 12.138 (0.012) | 0.5625604 (0.27)[2] | 744 (6) | 2459015.7317 (0.001)[2] |
| ATO J294.6544+39.5986 | 12.904 (0.012) | 3.605168 (402)[2] | 231 (83) | 2459017.7340 (0.014)[2] |
| ATO J293.4154+39.0755 | 14.941 (0.037) | 0.2503964 (1.3) | 95 (4) | 2459032.7268 (0.020) |
| ATO J294.7221+39.6485 | 15.439 (0.054) | 0.4078224 (5.7) | 121 (8) | 2459012.7394 (0.041) |
| ATO J293.2266+39.4496 | 16.197 (0.078) | 0.2527637 (1.0) | 447 (12) | 2459032.7962 (0.014) |
| ATO J293.1549+39.5200 | 16.605 (0.112) | 0.4003294 (3.1) | 541 (13) | 2459015.8301 (0.026) |
| ATO J294.3251+39.0326[1] | 16.888 (0.225) | 0.3772842 (860) | 618 (24) | 2459015.8333 (0.025) |
| ATO J293.7516+40.2587[1] | 17.091 (0.191) | 0.2793386 (230) | 693 (40) | 2459015.8301 (0.027) |

[1] Both of these faint stars only had usable data on two of our five nights. Hence, errors are orders of magnitude higher than in others for certain parameters, especially periods. As a result, periods derived by ATLAS were used in creating the phase plots for these two stars and not those listed in this table.

[2] The data collected by LO alone for these EA variables was not sufficient to determine a period or epoch. These values were determined after aligning with other data sources.

Table 2: Characteristics of ATLAS stars. Coordinates and magnitudes are from Gaia. Periods and amplitudes are from ATLAS (Heinze et al. 2018). Error estimates are not readily available for ATLAS parameters.

| Name | RA[deg][1] | DEC [deg][1] | G (error)[1] | Period [days] | Amplitude [mmag] | Type |
|---|---|---|---|---|---|---|
| KID 05015926[2] | 293.2068 | 40.1301 | 12.928 (0.003) | 0.362687 | 116 | CBH |
| V1130 Cyg[2] | 293.5137 | 39.7114 | 12.255 (0.002) | 0.562596 | 313 | CBF |
| ATO J294.6544+39.5986 | 294.6549 | 39.5987 | 12.963 (0.001) | 3.605378 | 186 | DBH |
| ATO J293.4154+39.0755 | 293.4154 | 39.0755 | 14.923 (0.003) | 0.250396 | 75 | NSINE |
| ATO J294.7221+39.6485 | 294.7222 | 39.6486 | 15.551 (0.004) | 0.407824 | 120 | SINE |
| ATO J293.2266+39.4496 | 293.2266 | 39.4496 | 16.200 (0.010) | 0.252765 | 448 | CBF |
| ATO J293.1549+39.5200 | 293.1550 | 39.5200 | 16.572 (0.012) | 0.400333 | 481 | CBH |
| ATO J294.3251+39.0326 | 294.3252 | 39.0327 | 16.950 (0.008) | 0.377943 | 613 | CBH |
| ATO J293.7516+40.2587 | 293.7516 | 40.2587 | 17.102 (0.009) | 0.279337 | 482 | CBF |

[1] From GAIA, J2000.0 (https://archives.esac.esa.int/gaia, see acknowledgments)
[2] These are the well-established variables used here for validation in Sec. 3.

Table 3: Terminology used during ATLAS classification (Heinze et al. 2018).

| ATLAS abbreviation | *Classification* |
|---|---|
| DBH | Detached eclipsing binary (ex. EA) |
| NSINE | Noisy sinusoidal variables such as spotted rotators, faint or low-amplitude delta Scuti stars and ellipsoidal variables (ex. ELL). |
| SINE | Sinusoidal variables. Ellipsoidal variables likely dominate this class. |
| CBF | Close binary, full period. The primary and secondary depths differ enough to fit the true period of the system. (ex. EW) |
| CBH | Close binary, half period. The primary and secondary depths are similar enough that the Fourier fit has overlapped transits to arrive at half the true period. (ex. EW) |

The performance of our absolute magnitude calibration and our amplitude determinations are explored in Fig. 5. Our derived magnitudes agree quite well with Gaia values and our error estimates are reasonable. The outliers are stars two and three (as ordered in Tabs. 1 and 2). These are both Algol type eclipsing binaries (EA). It is likely that the reported Gaia magnitude is a mean value and not the out-of-transit brightness that we report. Including in-transit magnitudes when finding the mean would underestimate the brightness in Gaia data, which is consistent with our difference. In comparing our derived amplitudes to those published by ATLAS (Heinze et al. 2018) we also see agreement except for two outliers. As will be discussed in Section 3, we believe that our results for star 2 (V1130 Cyg) are accurate. The other outlier, star 9, is the faintest in this discussion and we are very data limited. This is likely a genuine outlier in our processing. The LO photometry for these nine stars is available at https://www.dropbox.com/sh/p51g63yn8oh3m1p/AABTaVsgtGmBz3VsprBk903ca?dl=0

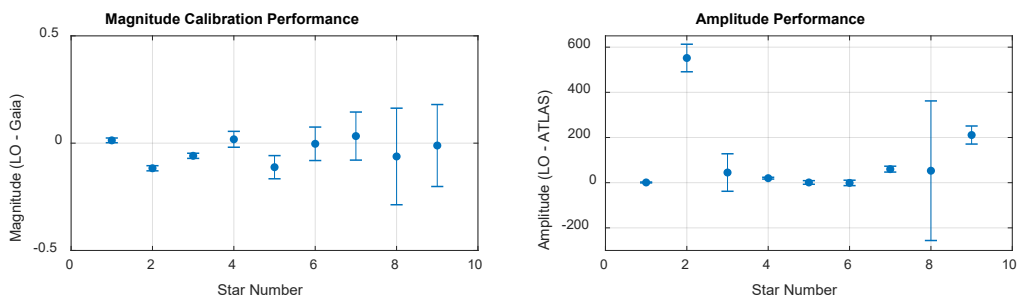

Figure 5: Performance of absolute magnitude calibrations and amplitude derivations. The graph on the left demonstrates that our calibrated magnitudes and error estimates are consistent with Gaia magnitudes. The two outliers are discussed in the text. The chart on the right demonstrates that our amplitude estimates are generally in agreement with those from ATLAS. The one extreme outlier is discussed in Section 3.

## 3 External Validation

We use photometry from several other surveys for comparison in this paper. The ATLAS survey is a very wide-field survey with the primary purpose of detecting hazardous asteroids approaching the Earth (Tonry et al. 2018). It has incidentally become a rich source of detections for other transient evens such as supernovae, flares, gamma-ray burst, and variable stars down to $V$~19. A dedicated list of new variable star candidates was released in 2018 (Heinze et al. 2018). As opposed to the high cadence of observations used during transit photometry, ATLAS generally includes a few photometric measurements of any given star each night. Variables are identified after calculating the Lomb-Scargle periodogram for each star and examining its false alarm probability (Heinze et al. 2018). While the ATLAS archive of new variables is significant, the faintness and low amplitude of these variables makes verification difficult. Seven of the variables detected by LO were previously only identified by ATLAS as candidates and thus do not appear in the American Association of Variable Star Observers (AAVSO) catalog. ATLAS light curves are publicly available via the Mikulski Archive for Space Telescopes website at http://archive.stsci.edu/.

The All-Sky Automated Survey for Supernovae (ASAS-SN) is an all-sky survey which collects photometry on stars down to $V$<17 every two to three days in the search for supernovae (Jayasinghe et al. 2018). The survey has incidentally identified over 66,000 variable stars and makes their light curves available to the public. Although observations are often spaced days apart, short-period variables are detectable thanks to an archive of observations extending back to 2014 and frequency analysis techniques very similar to those of ATLAS (Jayasinghe et al. 2018). ASAS-SN light curves are publicly available at https://asas-sn.osu.edu/.

Unlike ATLAS and ASAS-SN, the SuperWASP project aims at exoplanet discovery (Pollacco et al. 2006). It can survey the entire sky every 40 minutes, but at a somewhat poor accuracy of 1% for bright targets ($V$=7-11.5). Yet, its main goal is to make detections. And this level of accuracy is sufficient to detect large exoplanets, especially when its long baseline of observations allows for phase folding many periods to reduce noise. It is responsible for the discovery of nearly 200 exoplanets. SuperWASP light curves are available through the NASA Exoplanet Archive at https://exoplanetarchive.ipac.caltech.edu/.

The Kepler spacecraft collected photometry with the precision necessary to detect the transits of Earth-sized planets (Borucki et al. 2010). Its primary mission monitored a relatively small FOV in the region of Cygnus and Lyra. But the quality of the nearly continuous photometry of stars within this region is unprecedented. Kepler light curves are available through the NASA Exoplanet Archive at https://exoplanetarchive.ipac.caltech.edu/.

To gain confidence in our survey, we first compare our results for a known variable that has been observed by several surveys. The star KID 05015926 is a W Ursae Majoris-type eclipsing variable (EW) identified initially in Kepler data (Prsa et al. 2011). Light curves from Kepler, ASAS-SN, ATLAS, and SuperWASP are available for this star. It is $V$=12.952 and has a period of 0.362687 days (Prsa et al. 2011) with an amplitude slightly over 100 mmag based on multiple surveys. It has been observed by ATLAS 306 times over several years and was correctly classified as a contact or near-contact eclipsing binary. Photometry was obtained from CasJobs (see acknowledgments) and phase folded to the known period. Kepler and SuperWASP data were retrieved from the NASA Exoplanet Archive (see acknowledgments). ASAS-SN photometry was obtained directly from their online site (https://asas-sn.osu.edu/). The phase folded measurements from all these surveys are shown in Fig. 6. We bin data in each survey to have the same density in phase space as appears in the sparsest survey, ASAS-SN in this case with 200 observations. Kepler, ATLAS, and SuperWASP have 52,282, 306, and 1,689 observations, respectively. Binning improves the precision of each resulting measurement by reducing Gaussian white noise and allows for a fair quality comparison between sparse surveys and those having an abundance of data. Magnitudes are shown relative to the mean for simplicity since these surveys have a variety of passbands and zero points. The detections are clear and consistent among all surveys.

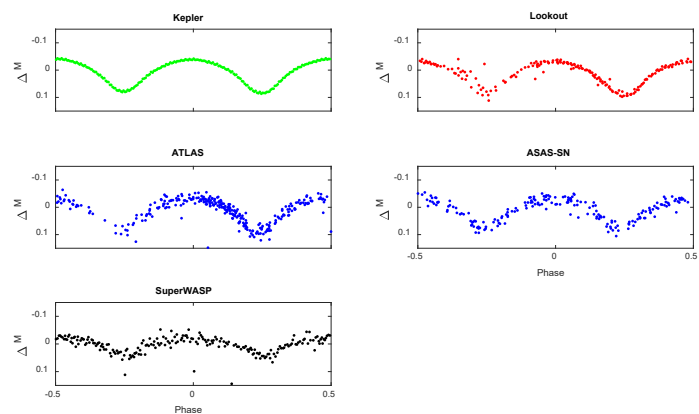

Figure 6. Phase folded light curves for KID 05015926 with all data binned to a common sampling to fairly compare surveys having wide ranges of observations. LO results compare well with other ground-based surveys.

Next, we test our detection of an eclipsing binary of Algol type (EA). The star V1130 Cyg is *V*=12.8 and has an amplitude of nearly one magnitude with a period of 0.562596 days (Miller 1966). Light curves from these surveys are shown in Fig 7. Although LO results are poorly sampled during the secondary transit, the quality of our photometry compares favorably with other surveys, excluding Kepler. Our calculated transit depth of 744 (6) mmag differs greatly from the ATLAS value of 313 mmag. A visual examination of the phase curves from all five surveys (including ATLAS), however, confirms that the larger value is correct. A duration of 172.53 (1.8) minutes was determined after fitting the data using the model of Mandel and Agol (Mandel et al. 2002). This model is strictly speaking designed for exoplanet light curve fitting and several of the derived parameters will not apply to an eclipsing binary. The duration, however, is sensibly fit.

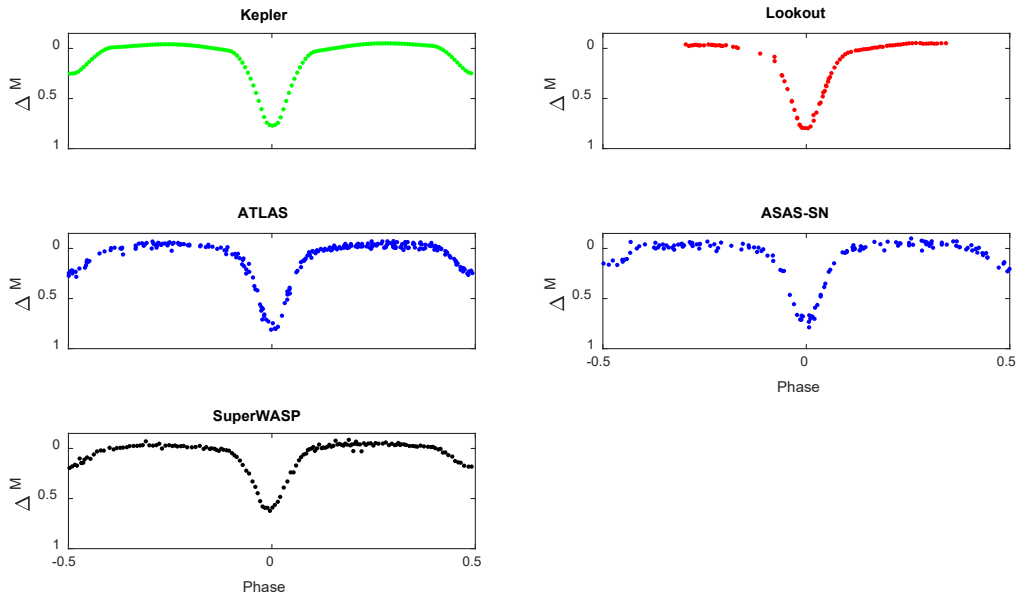

Figure 7. Phase folded light curves for V1130 Cyg with all data binned to a common sampling. Although the secondary transit was not observed by LO during its five sessions, the quality compares favorably with other surveys. It is based on observations from only five nights.

## 4 Results

Having some confidence in the LO results, we now compare light curves for stars which have only been identified as variable in the ATLAS and LO surveys. All the light curves are provided in Fig. 8-14. Our data is binned to have approximately the same phase density as ATLAS results.

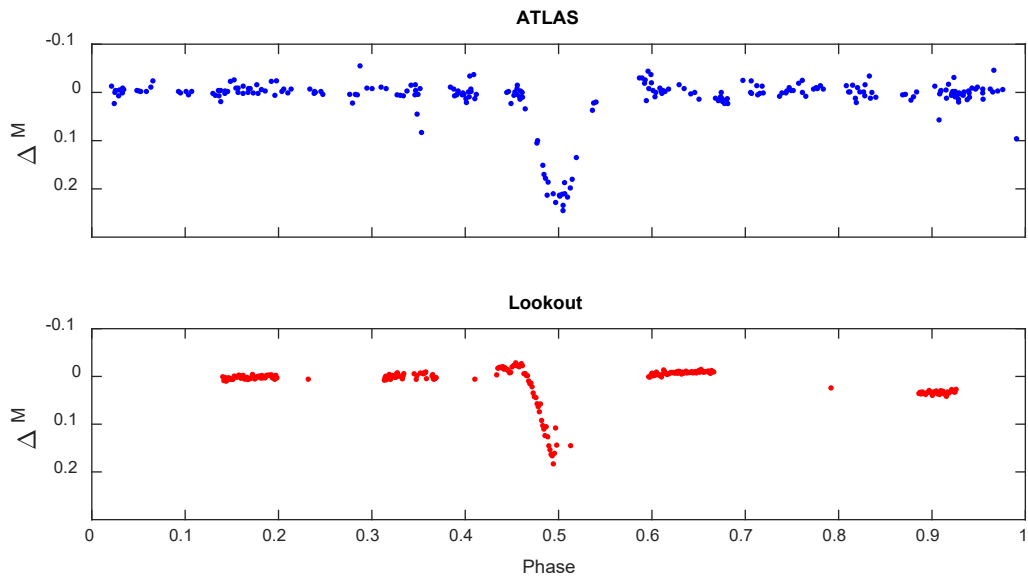

Figure 8: ATLAS and LO light curves for ATO J294.6544+39.5986 phase folded to a period of 3.605378 days. Although phase coverage is limited in the LO photometry, the nature of the star as a detached eclipsing binary is confirmed.

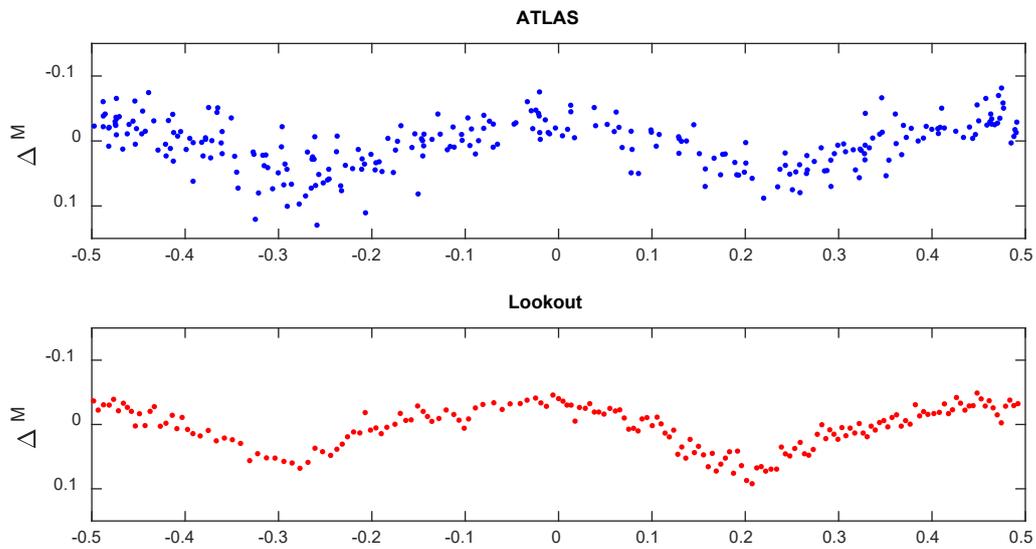

Figure 9: ATLAS and LO light curves for ATO J293.4154+39.0755 phase folded to a period of 0.2503964 days. The variable nature of this variable is confirmed.

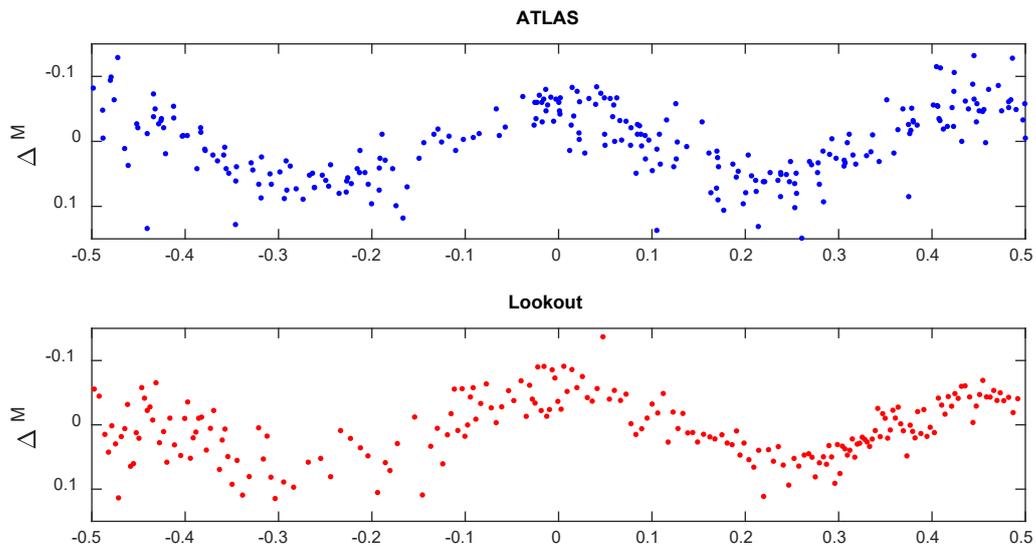

Figure 10: ATLAS and LO light curves for ATO J294.7221+39.6485 phase folded to a period of 0.4078224 days. The variable nature of this variable is confirmed.

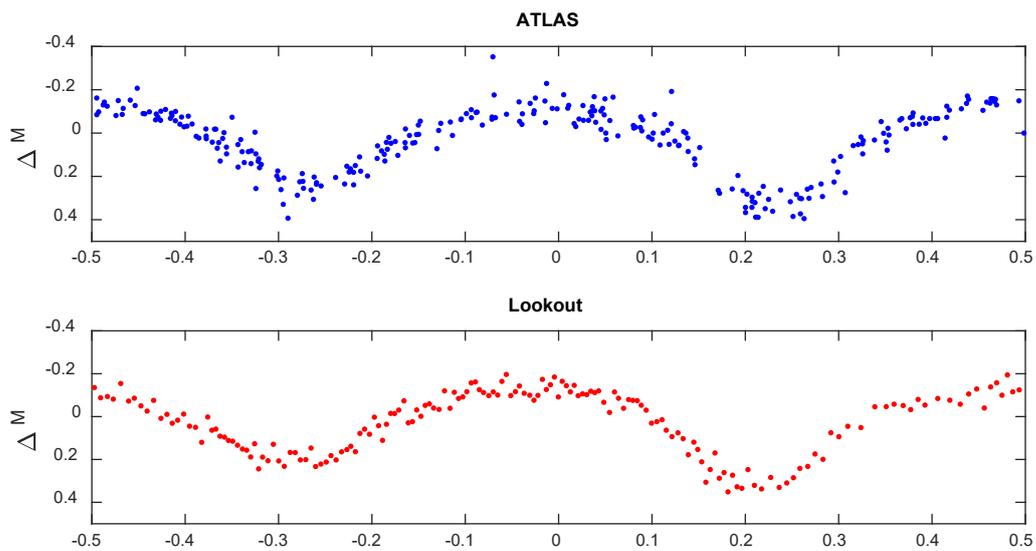

Figure 11: ATLAS and LO light curves for ATO J293.2266+39.4496 phase folded to a period of 0.2527637 *days*. The close binary nature of this variable is confirmed and there is a clear difference between primary and secondary eclipse depths.

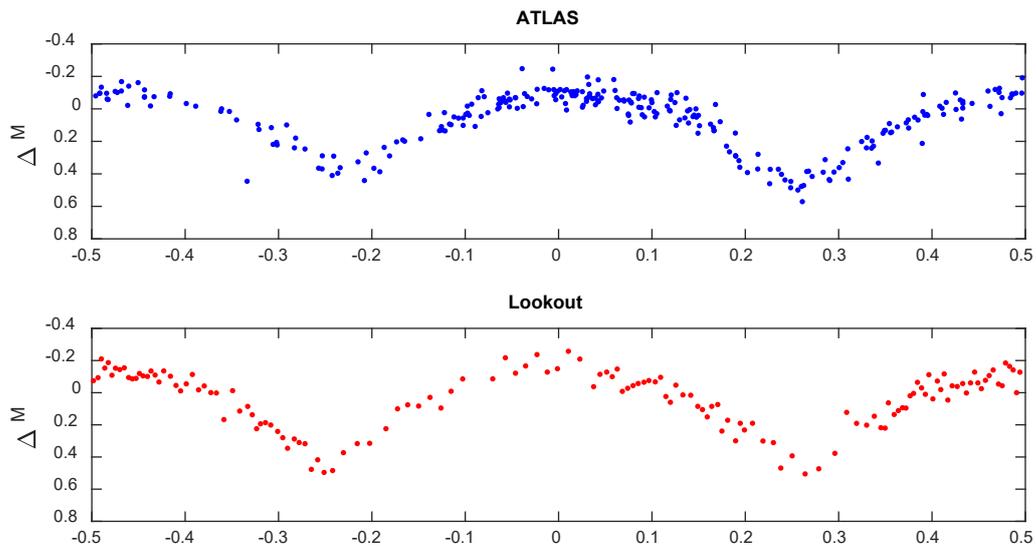

Figure 12: ATLAS and LO light curves for ATO J293.1549+39.5200 phase folded to a period of 0.4003294 days. The light curves are consistent with ATLAS's classification of this system as a close binary.

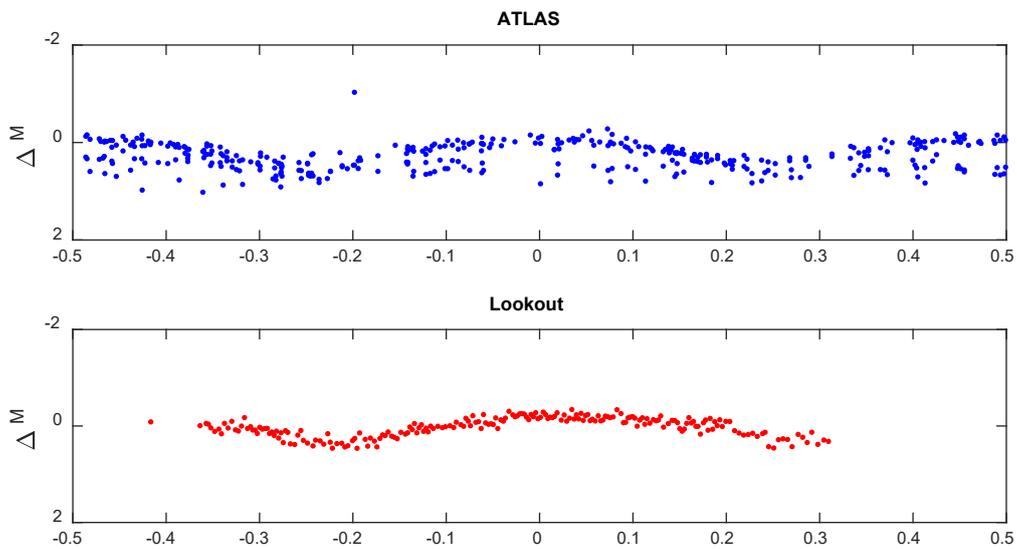

Figure 13: ATLAS and LO light curves for ATO J294.3251+39.0326 phase folded to a period of 0.377943 days. Although both results are noisy and LO results do not provide full phase coverage, the variable nature of this star is confirmed.

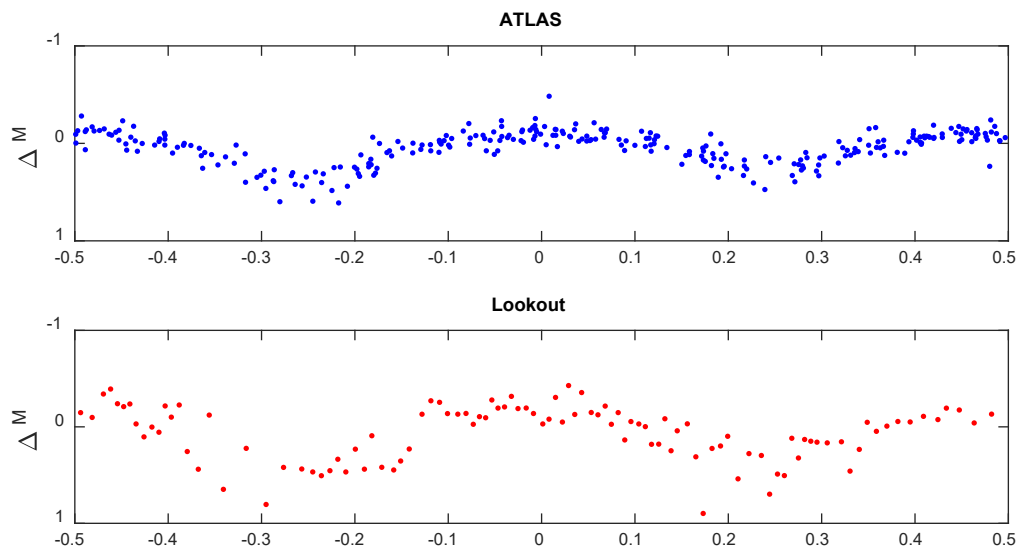

Figure 14: ATLAS and LO light curves for ATO J293.7516+40.2587 phase folded to a period of 0.279337 days. Results are consistent with ATLAS's classification as a close binary.

## 5　Conclusions

The Lookout Observatory has finished its commissioning phase and demonstrated its photometric goals by automatically observing short-period variables at a quality that is comparable to other ground-based programs. Seven candidate variables identified by the ATLAS survey have been independently confirmed. Many FOVs have been observed over the past two years, often multiple times. This survey will continue to focus on exoplanet observations, but variable star data will be released as it becomes available.

Acknowledgements: In addition to the support of the Astronautical Engineering Department at the US Air Force Academy (specifically Col Luke Sauter for the powerful gift of time), the authors would like to thank the LO construction team, Savannah Jane and P.P. PA#: USAFA-DF-2020-325

CasJobs was used in retrieving ATLAS light curves as authored by the JHU/SDSS team. http://casjobs.sdss.org/CasJobs.

This research has made use of the NASA Exoplanet Archive, which is operated by the California Institute of Technology, under contract with the National Aeronautics and Space Administration under the Exoplanet Exploration Program.

This paper makes use of data from the first public release of the WASP dataset provided by the WASP consortium and services at the NASA Exoplanet Archive, which is operated by the California Institute of Technology, under contract with the National Aeronautics and Space Administration under the Exoplanet Exploration Program.

We acknowledge with thanks the variable star observations from the *AAVSO International Database* contributed by observers worldwide and used in this research.

This work presents results from the European Space Agency (ESA) space mission Gaia. Gaia data are being processed by the Gaia Data Processing and Analysis Consortium (DPAC). Funding for the DPAC is provided by national institutions, in particular the institutions participating in the Gaia MultiLateral Agreement (MLA). The Gaia mission website is https://www.cosmos.esa.int/gaia. The Gaia archive website is https://archives.esac.esa.int/gaia.